\documentclass[twoside]{article}
\usepackage{qic,epsfig,rotating}
\usepackage{graphicx,epstopdf}

\newcommand{\figref}[1]{Fig.~\ref{fig:#1}}
\newcommand{\tabref}[1]{Table~\ref{tab:#1}}
\newcommand{\secref}[1]{Sec.~\ref{sec:#1}}

\begin{document}
\setlength{\textheight}{8.0truein}    

\runninghead{Demonstration of a scalable, multiplexed ion trap for quantum information processing}{D.~R.~Leibrandt et al.}

\normalsize\textlineskip
\thispagestyle{empty}
\setcounter{page}{1}

\copyrightheading{0}{0}{2003}{000--000}

\vspace*{0.88truein}

\alphfootnote

\fpage{1}

\centerline{\bf
DEMONSTRATION OF A SCALABLE, MULTIPLEXED ION TRAP}
\vspace*{0.035truein}
\centerline{\bf FOR QUANTUM INFORMATION PROCESSING}
\vspace*{0.37truein}
\centerline{\footnotesize
D.~R.~LEIBRANDT, J.~LABAZIEWICZ, R.~J.~CLARK, and I.~L.~CHUANG}
\vspace*{0.015truein}
\centerline{\footnotesize\it Center for Ultracold Atoms, Research Laboratory of Electronics, \& Department of Physics}
\baselineskip=10pt
\centerline{\footnotesize\it Massachusetts Institute of Technology}
\baselineskip=10pt
\centerline{\footnotesize\it Cambridge, Massachusetts 02139, USA}
\vspace*{10pt}
\centerline{\footnotesize 
R.~J.~EPSTEIN, C.~OSPELKAUS, J.~H.~WESENBERG, J.~J.~BOLLINGER, D.~LEIBFRIED, and D.~J.~WINELAND}
\vspace*{0.015truein}
\centerline{\footnotesize\it National Institute of Standards and Technology}
\baselineskip=10pt
\centerline{\footnotesize\it Boulder, Colorado 80305, USA}
\vspace*{10pt}
\centerline{\footnotesize
D.~STICK, J.~STERK, and C.~MONROE}
\vspace*{0.015truein}
\centerline{\footnotesize\it University of Michigan}
\baselineskip=10pt
\centerline{\footnotesize\it Ann Arbor, Michigan 48109, USA}
\vspace*{10pt}
\centerline{\footnotesize
C.-S.~PAI, Y.~LOW, and R.~FRAHM}
\vspace*{0.015truein}
\centerline{\footnotesize\it Bell Laboratories, Alcatel-Lucent}
\baselineskip=10pt
\centerline{\footnotesize\it Murray Hill, New Jersey 07974, USA}
\vspace*{10pt}
\centerline{\footnotesize 
R.~E.~SLUSHER}
\vspace*{0.015truein}
\centerline{\footnotesize\it Georgia Tech Research Institute, Georgia Institute of Technology}
\baselineskip=10pt
\centerline{\footnotesize\it Atlanta, Georgia 30332, USA}
\vspace*{0.225truein}
\publisher{(received date)}{(revised date)}

\vspace*{0.21truein}

\abstracts{
A scalable, multiplexed ion trap for quantum information processing is fabricated and tested.  The trap design and fabrication process are optimized for scalability to small trap size and large numbers of interconnected traps, and for integration of control electronics and optics.  Multiple traps with similar designs are tested with $^{111}$Cd$^+$, $^{25}$Mg$^+$, and $^{88}$Sr$^{+}$ ions at room temperature and with $^{88}$Sr$^+$ at 6 K, with respective ion lifetimes of 90 s, 300 $\pm$ 30 s, 56 $\pm$ 6 s, and 4.5 $\pm$ 1.1 hours. The motional heating rate for $^{25}$Mg$^{+}$ at room temperature and a trap frequency of 1.6 MHz is measured to be 7 $\pm$ 3 quanta per millisecond. For $^{88}$Sr$^{+}$ at 6 K and 540 kHz the heating rate is measured to be 220 $\pm$ 30 quanta per second.
}{}{}

\vspace*{10pt}

\vspace*{3pt}
\communicate{to be filled by the Editorial}

\vspace*{1pt}\textlineskip    

\section{Introduction}

Trapped ions show much promise for achieving large-scale quantum information processing (QIP).  One proposed architecture uses electronic or nuclear states of ions trapped in an RF Paul trap to store information, and laser-ion interactions to accomplish initialization, logic gates, and readout.  Two-qubit gates are mediated by the shared motional mode of two ions in a single trap, and ions are shuttled between zones in an array of ion traps to perform two-qubit gates between arbitrary pairs of ions \cite{wineland98, cirac00, kielpinski02}.  Experiments have demonstrated movement of ions in multiplexed ion trap arrays \cite{rowe02, hensinger06} and implemented simple algorithms \cite{riebe04, barrett04, chiaverini04, haffner05}.  The current challenge for this architecture of trapped ion QIP is integration and scaling to large numbers of ions.

This work demonstrates a scalable, multiplexed ion trap designed for large-scale quantum information processing \cite{kim05}.  The aluminum trap electrodes are fabricated on a silicon substrate using standard VLSI (very-large-scale integration) techniques.  All of the processing is done below 400$^{\circ}$ C, so future versions of the trap can be built directly on top of CMOS (complementary metal-oxide-semiconductor) control electronics.  The trap is tested both at room temperature and at 6 K for improved ion lifetime and heating rate.  The ion lifetime and the heating rate of the ion motional state are measured to determine if the trap is suitable for QIP.

This paper proceeds as follows.  \secref{trapDesign} describes the trap design and fabrication process.  Secs.~\ref{sec:trapTestingUM}, \ref{sec:trapTestingNIST}, and \ref{sec:trapTestingMIT} describe trap operation and testing with $^{111}$Cd$^+$, $^{25}$Mg$^+$, and $^{88}$Sr$^{+}$ ions at the University of Michigan, the National Institute of Standards and Technology, and the Massachusetts Institute of Technology, respectively.  Finally, \secref{conclusion} summarizes and concludes.

\section{Trap design, fabrication, and packaging}
\label{sec:trapDesign}

The ion trap is a surface-electrode RF Paul trap \cite{chiaverini05, pearson06, seidelin06} fabricated on a silicon substrate \cite{kim05, brownnutt06, britton06, stick06}.  The basic idea is to use materials and processes that are as close to the standard silicon VLSI processing techniques as possible so that scaling to large ion numbers can be easily achieved. This restricts one to using materials that are allowed in a silicon processing clean room. For example, gold and silver are not allowed in the initial fabrication although they can be added in a separate processing step. The initial traps are fabricated using aluminum metal, a common VLSI material with low resistivity. A thermal oxide forms at the Al surface which is known to charge upon UV irradiation \cite{rageh77}. If this were to degrade the trap performance, a W or Ti/W layer of the order of 60 $\mu$m thick could be deposited on top of the Al electrodes in order to improve the metal surface for ion trapping (tungsten has been used successfully for previous micron scale ion traps \cite{deslauriers06}); however, this was not attempted in the early experiments described here.

\begin{figure}
\centering
\includegraphics[width=12cm]{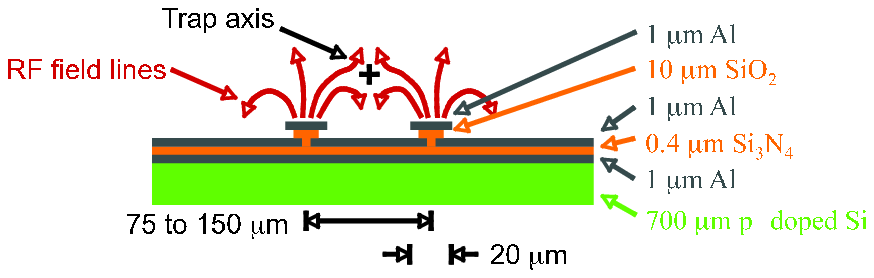} 
\fcaption{\label{fig:layers}
Cross section of the surface-electrode ion trap.  DC voltages are applied to the three electrodes in the middle layer of aluminum and RF is applied to the two raised rails in the top layer of aluminum.  Ions are trapped at the RF quadrupole null located 39 to 79 $\mu$m above the surface of the DC electrodes, indicated by a black cross.  Not to scale.}
\end{figure}

\begin{figure}
\centering
\includegraphics[width=14cm]{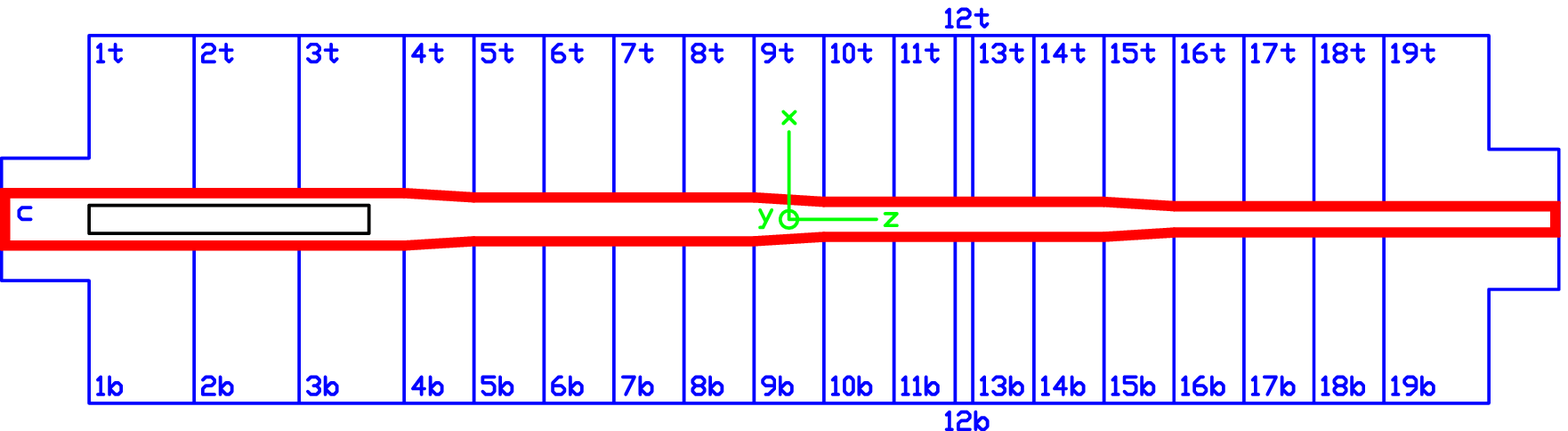} 
\fcaption{\label{fig:layout}
Layout of the surface-electrode ion trap.  The DC electrodes are outlined and labeled in blue and the RF electrodes are drawn in solid red.  The black rectangle is an optional through wafer slot used for loading.  The DC electrodes are between 50 $\mu$m and 300 $\mu$m wide along the trap axis.  The RF electrodes are 20 $\mu$m wide and are separated by 150 $\mu$m, 125 $\mu$m, 100 $\mu$m, and 75 $\mu$m center-to-center in the four regions going from left to right.}
\end{figure}

A cross section of the trap is shown in \figref{layers} and the layout of the ion trap chip is shown in \figref{layout}. The substrate is a highly doped conductive silicon wafer with a resistivity of 0.018 $\Omega$ cm. Even though the silicon has a much higher resistivity than the metal, its thickness is 700 $\mu$m and the resulting resistance is similar to that of the 1 $\mu$m thick metal layers. However, the doped Si is very lossy at the RF frequencies used for trapping so it can not be exposed to the RF trapping fields.

The first layer of aluminum serves as a ground plane to isolate the trap electrodes from the lossy Si substrate. This insures that the Q of the RF circuit is sufficiently high for efficient RF drive. The ground plane is 1 $\mu$m thick Al with a resistance of 0.027 $\Omega$/\square\fnm{a}\fnt{a}{The unit $\Omega$/\square~is ohms per square, which is the resistance of a square area of the metal film.}. The second metal layer comprising the DC control electrodes is insulated from the ground plane by a 0.4 $\mu$m layer of silicon nitride with a dielectric constant of 7.5. In some of the traps the ground plane is omitted. In this case the nitride layer is deposited directly on the Si substrate and the DC control electrode metallic layer isolates the RF electrodes from the Si substrate. The DC control electrode structures are over-coated by a 10 $\mu$m thick dielectric layer of PETEOS (plasma-enhanced tetraethylorthosilicate), which is a stress balanced SiO$_2$ with a dielectric constant of 4. The silicon dioxide is made as thick as possible to reduce the capacitance and increase the breakdown voltage between the DC and RF electrodes \cite{kim05, leibrandt07a}. The capacitance must be minimized since the dissipated RF power, which can easily limit the voltage that can be applied to the trap, varies as the square of the capacitance. It has been found that this SiO$_2$ layer must be deposited with carefully controlled temperature, e.g. deposition of a micron or two and then a pause for thermal equilibration. It is possible that this deposition process could result in micro-voids that slowly outgas and cause ion de-trapping via collisions. The third metal layer comprising the RF electrodes is also 1 $\mu$m of Al. There is an effective capacitance divider formed by the relatively high capacitance of the control electrodes to ground (via the thin nitride insulating layer with a high dielectric constant) and the much smaller (roughly one fiftieth) capacitance between the DC control electrodes and the RF rails. This capacitive divider effectively grounds the RF potentials that can be picked up by the close proximity of the RF and DC control electrodes. Additional capacitors are connected to the DC control electrodes on filter circuits outside the vacuum chamber to further RF ground the control electrodes.

A slot for loading or vertical laser access was added to some of the fabricated traps. The loading slot is an advantage for avoiding contamination of the electrode surfaces from the neutral-atom oven used to load the trap. This oven is placed behind the trap chip and atoms emerging through the slot are photoionized. This through slot was fabricated by etching from the back side of the wafer. First the wafer was sent out for grinding down to a thickness of 400 $\mu$m. It was then chemically mechanically polished (CMP) using standard techniques. Next a mask was used to etch through the wafer from the back side. This dry etching process resulted in very straight slot walls with less than a 5\% slant toward the surface. The walls of the slot are quite rough compared to the other surfaces in the trap (1--5 $\mu$m peak-to-peak roughness).

\begin{figure}
\centering
\includegraphics[width=8cm]{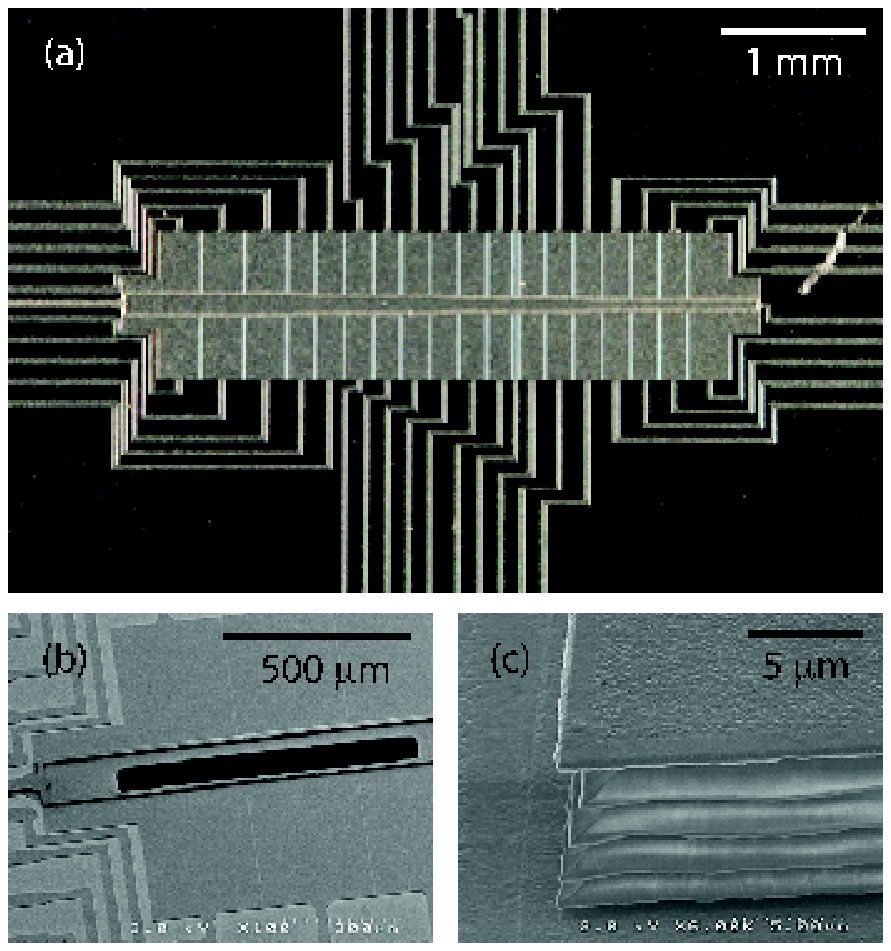} 
\fcaption{\label{fig:SEMpics}
Pictures of the fabricated trap structures.  (a) Optical picture of the trap layout.  This particular trap does not have a through wafer loading slot.  (b) SEM (scanning electron microscope) picture of the through wafer loading slot located in the center electrode in the trap region with the largest RF rail separation.  (c) SEM picture of the side of the 10 $\mu$m SiO$_2$ layer used to raise the RF electrodes above the DC electrodes.  The SiO$_2$ is striated because it is deposited in several layers for stress relief.}
\end{figure}

Pictures of the fabricated trap structures are shown in \figref{SEMpics}. Ions are confined to the RF quadrupole null located 39 to 79 $\mu$m (depending on the separation of the RF rails) above the surface of the chip by an RF voltage between 100 and 370 V amplitude at roughly 40 MHz applied to the two RF rails.  The RF electrodes are able to withstand up to 400 V amplitude before electrical arcing destroys the trap.  Confinement along the trap axis is accomplished by DC voltages between -15 and 15 V applied to the 19 pairs of DC control electrodes on the outside of the RF rails and to the center electrode in between the RF rails (as shown in \figref{layout}).  Note that the metal electrodes' surfaces are very smooth ($\sim$ 100 nm peak-to-peak roughness) and that the electrode edges are very well defined (300--500 nm peak-to-peak roughness). In fact, the very sharp electrode edges could present a problem if field emission occurs at high RF voltages.

The silicon trap chips were mounted in 100-pin ceramic pin grid array (CPGA) carriers (Global Chip Materials PGA10047002). Contacts to the electrodes are made to 300 $\times$ 300 $\mu$m$^2$ Al pads by wire bonding at the edge of the chip. A spacer was required between the trap chip and the carrier in order to elevate the chip above the upper surface of the carrier for laser access across the top of the trap chip. This spacer was made from 1.27 mm thick 96\% alumina. A 2 mm $\times$ 10 mm slot was laser machined into the alumina spacer for access to the ion loading slot. 

Attaching the chip to the spacer, and the spacer to the carrier was accomplished by epoxy (EPO TEK 353), ceramic paste, or soldering. Epoxy is the simplest method, however there are concerns about critical baking procedures and contributions to the base vapor pressure in the UHV (ultra-high vacuum) chamber. Epoxy also has the disadvantage that it can not be cleaned just prior to mounting the carrier and chip in the vacuum chamber since residue from the partially soluble epoxy could coat the trap electrode surfaces.

Ceramic paste is another method for mounting the chip and spacer. The ceramic is not a good ``glue'' and a slight overlap of the ceramic paste around the edge of the chip or spacer is required. The ceramic paste sometimes does not adhere well to the chip or spacer resulting in the chip detaching during wire bonding. Another problem with the ceramic paste is that, as was the case with epoxy, the ion trap can not be cleaned after mounting in the CPGA because the ceramic paste is partially soluble.

\begin{figure}
\centering
\includegraphics[width=8cm]{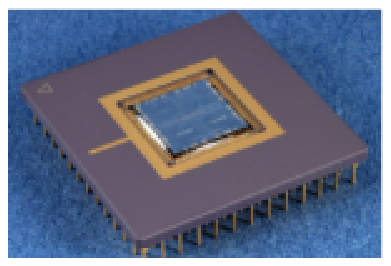} 
\fcaption{\label{fig:packagedTrap}
Packaged ion trap.  The ion trap is packaged and wire bonded in a 100-pin ceramic pin grid array.  This trap chip is mounted with ceramic paste that can be seen around the edges of the chip.  The ion trap chip is 1.0 cm by 1.0 cm.}
\end{figure}

The best method found for mounting the chip and spacer is by solder contact. This requires that the chip and spacer surfaces be coated with gold. Thin strips of 80/20 Au/Sn are placed between the two gold coated surfaces to be bonded. The assembly is heated to approximately 285 C in order to melt the Au/Sn alloy. The hot Sn diffuses into the gold bonding surfaces leaving a much higher melting temperature alloy in the bonding region. In this manner a second bond can be made between the trap chip and the spacer without unbonding the carrier/spacer bond. The solder bonds did not cause any fracturing of the trap structures even though the coefficients of thermal expansion are different by a factor of two for the ceramic and the silicon chip. This solder bond is strong and clean. It can be re-cleaned after testing the chip just prior to installing into the vacuum chamber.  A packaged ion trap is shown in \figref{packagedTrap}.

Ion traps for QIP must be able to store ions reliably for the duration of the computation and have an ion motional heating rate much smaller than the speed of the two qubit gates.  The next three sections describe measurements of the ion lifetime and motional heating rate at the University of Michigan, the National Institute of Standards and Technology, and the Massachusetts Institute of Technology.  The University of Oxford and the University of Innsbruck also received traps for testing, but were not able to trap ions due to shorting between the electrodes \cite{OxfordSMITtesting,InnsbruckSMITtesting}.

\section{Trap testing at the University of Michigan}
\label{sec:trapTestingUM}

The time-averaged pseudopotential of the ion trap is an anisotropic three-dimensional harmonic oscillator potential.  QIP experiments require that the ion is cooled to near the ground motional state in all three dimensions.  One disadvantage of the surface-electrode geometry is that the cooling lasers are constricted to a plane parallel to the chip.  Because laser cooling cools only the modes of the ion motion whose principal axes have a component along the laser wavevector(s), the DC voltages must be chosen such that none of the principal axes of the pseudopotential are orthogonal to the chip surface and therefore the cooling laser(s).  Such voltages were found by numerically modeling the potentials above the trap electrodes using a boundary element electrostatics solver \cite{CPO}.  The voltages used for trap testing with $^{111}$Cd$^+$ at the University of Michigan are given in \tabref{UMvoltages}.  With this set of voltages, the principal axis coordinate system is rotated 12 degrees about the z-axis with respect to the coordinate system shown in \figref{layout}.  The calculated secular frequencies and trap depth are $(\omega_x', \omega_y', \omega_z) / (2 \pi) = (5.7, 5.9, 0.5)$ MHz and 0.5 eV.

\begin{table}
\centerline{
\begin{tabular}{ll}
Electrode		& Voltage \\
\hline
RF			& 370 V at 32 MHz plus 2.34 V DC \\
C			& 2.34 V \\
2T			& 3.12 V \\
2B			& 6.12 V \\
3T			& 1.90 V \\
3B			& 1.22 V \\
4T			& 3.12 V \\
4B			& 6.12 V
\end{tabular}
}
\tcaption{\label{tab:UMvoltages}
Typical voltages used for trap testing at the University of Michigan.  The electrodes not listed are grounded.  These voltages trap the ion between electrodes 3B and 3T.}
\end{table}

The University of Michigan group received three electrically verified ion traps. Of these, ions were trapped in only the first trap, which did not have an aluminum ground plane.  This resulted in a lower Q ($\sim$ 100) than the traps with the aluminum ground plane due to RF losses in the silicon substrate. Ions were loaded by photoionization using a pulsed 229 nm laser (one laser, two photon transition), and Doppler cooled on the $^{2}$S$_{1/2} \left| F=1,m_F=-1 \right>$ $\leftrightarrow$ $^{2}$P$_{3/2} \left|F=2,m_F=-2\right>$ transition at 280 nm (up to 5 mW in a 20 $\mu$m radius beam, red detuned by 30--100 MHz from the $^{111}$Cd$^+$ transition) with a repumper addressing the  $^{2}$S$_{1/2} \left| F=0,m_F=0 \right>$ $\leftrightarrow$ $^{2}$P$_{3/2} \left|F=1,m_F=-1\right>$ transition.  The Doppler cooling lasers were oriented at 45 deg to the trap axis.  In this first trap ions were trapped in the region with the 150 $\mu$m RF electrode spacing and the height of the ion 79 $\mu$m above the surface.  Ions were trapped using the DC voltages given in \tabref{UMvoltages} and between 250 and 370 V RF.  At 370 V RF the mean ion lifetime was about 90 s, with a maximum ion lifetime of about 5 minutes.  The low lifetime is attributed to background gas collisions; the pressure of the chamber was measured to be $2.0 \times 10^{-8}$ Pa at a distance far from the ion trap, but outgassing of the chip/epoxy/socket would have increased the local pressure at the trap above this measured chamber base pressure. With the photoionization laser on, the ion position drifted in from the region closer to where the RF wires attach to the rails. This is attributed to a background electric field associated with the pulsed ionization laser, since once that was turned off the ion stayed in its expected region (defined by the applied control voltages). By simply blocking the cooling beam it was determined that the ion would last at most 5 s without cooling before it left the trap.  Over the course of a few weeks of trapping, this first trap developed electrical shorts between several of the DC electrodes and the silicon substrate.

The next trap had a backside loading slot, though this is not necessary to trap cadmium (which has a relatively high room temperature vapor pressure, eliminating the need for line of site access from the oven to the trapping region). In fact, it was determined that the loading slot can be detrimental since the central DC electrode which was used to compensate for the vertical electric field due to the other DC electrodes was now absent. The aluminum ground plane resulted in a higher Q ($\sim$ 150) which resulted in less power dissipation due to the RF drive. No ions, however, were trapped in this trap.

The third trap did not have a backside loading slot and did have an aluminum ground plane. Again, no ions were trapped in this trap.  In situ measurements of the electrical properties looked good (Q $\sim$ 150, RF and DC electrode capacitances normal).

\section{Trap testing at NIST}
\label{sec:trapTestingNIST}

Trap testing at NIST was performed with $^{25}$Mg$^{+}$. Neutral magnesium was photoionized in the trap through a two-photon  process using 1--10 mW of CW 285 nm laser power focused to a waist of about 30 $\mu$m. The ions were excited on the closed $^2$S$_{1/2} \left| F=3,m_F=-3 \right>$ $\leftrightarrow$ $^{2}$P$_{3/2} \left|F=4,m_F=-4\right>$ transition at 280 nm with a circularly polarized laser beam that was aligned with a 1 mT quantizing magnetic field and detuned -20 MHz from the transition center frequency for optimal Doppler cooling. Due to imperfect cycling and the hyperfine structure of $^{25}$Mg$^+$, a second resonant laser beam was used to repump out of the F=2 ground-state manifold. For the heating rate measurements described below, intensities near saturation were used \cite{epstein07}.  Single ions were loaded with the cooling beam momentarily detuned by about -300 MHz in the center of the loading slot (between electrodes 2T and 2B) at a calculated well depth of about 0.050 eV and calculated trap frequencies of approximately (5.0, 5.4) MHz radial and 1 MHz axial. The principal axes had an angle of about (79, 47) degrees (radial directions) and 45 degrees (axial direction) to the wavevector of the cooling beam. Micromotion compensation had to be frequently performed in the load zone with large variations of the compensation voltages for different ion loads. This was attributed primarily to charging produced by laser beam scatter on the electrode structures. For this reason cooling laser intensities were kept as small as possible for loading ($\sim$ 20 $\mu$W in a 30 $\mu$m waist). After loading, the ions were transported to the trap zone between electrodes 5T and 5B, where they resided at a distance of about 60 $\mu$m from the nearest trap electrode. The calculated trap frequencies in this location were (7.3, 7.7) MHz (radial directions) and 1.6 MHz (axial direction) with a trap depth calculated to be approximately 0.064 eV.

Three different versions of the basic trap design where tested at NIST. The first chip was attached with EPO TEK 353\fnm{b}\fnt{b}{Commercial products are mentioned for the sake of technical completeness; this does not imply endorsement by NIST.} epoxy to the chip carrier. Despite a careful bake-out and leak testing it was only possible to reach a base pressure of $6.7 \times10^{-8}$ Pa in the vacuum enclosure with this chip. Ion loading was not successful, possibly because ion lifetimes in the trap were limited to very short durations due to the relatively high background pressure (fluorescence was registered with an electron-multiplication CCD camera).

The second trap was mounted in the chip carrier using ceramic paste. After baking, a base pressure of $2.7 \times 10^{-8}$ Pa was achieved.  Subsequent to bakeout, some electrodes exhibited fluctuating resistances to ground in the range between 100 $\Omega$ and 100 k$\Omega$. The resistance depended on the polarity of the applied potential, indicating diode-like behavior. Fortunately, none of the electrodes adjacent to the loading zone exhibited this resistance and backside loading of single ions into the load zone of the chip was successful. The trap lifetimes were between 30 s and 2 min. Possibly due to large and variable stray fields it was also hard to reproduce the loading position from load to load. In the process of improving loading conditions one of the control electrodes in the load zone developed a short to ground after a potential of approximately 10 V was applied to it continuously for one week. Therefore further trapping had to be abandoned with this chip. Subsequent breakdown tests with other DC electrodes on the same chip revealed a diode-like behavior of the current vs.~voltage with a fast breakdown (order a few seconds) with approximately 30 V applied.  This same mechanism may have been responsible for the shorting of the load zone electrode.

With the third chip tested, a base pressure of $6.7 \times 10^{-9}$ Pa was achieved after bake out. No electrode corruption was detected over the several weeks of runnning during which ions were routinely loaded. Ion lifetimes were approximately 5 minutes in the presence of the cooling laser and 5--10 s without cooling light. For other surface-electrode traps at the same base pressure the ion lifetime is observed to be 20 minutes to several hours \cite{seidelin06}.  Single ions were transported from the load zone to the trap zone between electrodes 5T and 5B for performing heating rate measurements. These measurements were based on a simplified technique utilizing Doppler-recooling as described in \cite{epstein07}. The heating rate was 7$\pm$3 quanta/ms at a distance of about 60 $\mu$m from the nearest electrode and a trap frequency of 1.6 MHz.

\section{Trap testing at MIT}
\label{sec:trapTestingMIT}

\begin{figure}
\centering
\includegraphics[width=8cm]{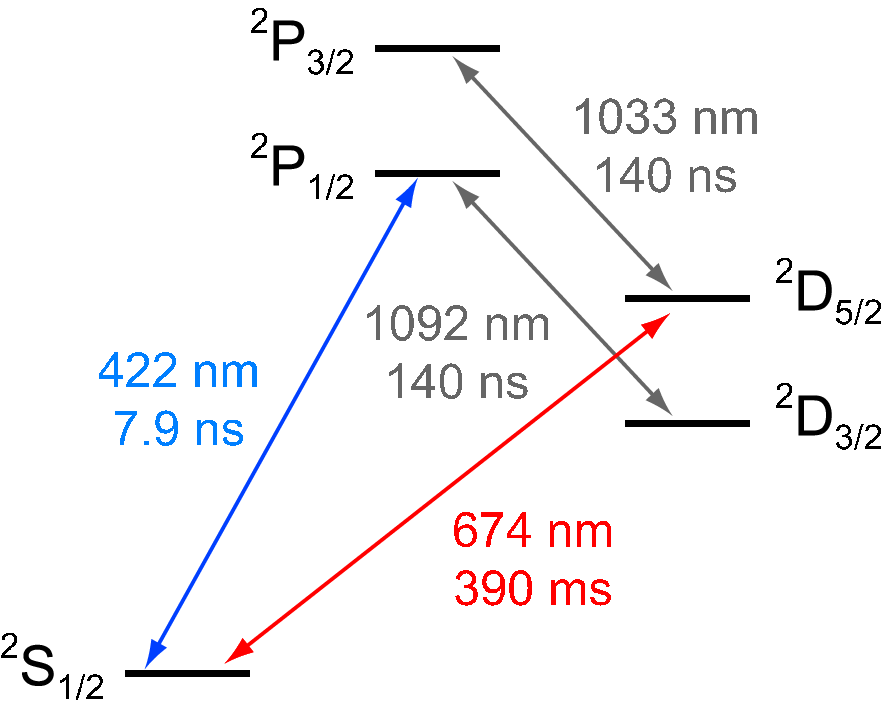} 
\fcaption{\label{fig:levelDiagram}
Level diagram of $^{88}$Sr$^+$.  Dopper cooling is performed on the 422 nm $S_{1/2} \leftrightarrow P_{1/2}$ transition with an additional 1092 nm laser to repump the $D_{3/2}$ state.  Resolved sideband cooling and temperature measurement are performed on the 674 nm $S_{1/2} \leftrightarrow D_{5/2}$ transition.  A 1033 nm laser is used to repump the $D_{5/2}$ state between experiments.}
\end{figure}

Trap testing at MIT was performed with $^{88}$Sr$^+$.  The ions are loaded into the trap by photoionization of a thermal atomic beam as it passes through the trap region.  Photoionization proceeds via a resonant two-step process with 460 nm and 405 nm lasers (600 $\mu$W of each in a 50 $\mu$m radius beam) \cite{vant06, brownnutt07}.  A level diagram of $^{88}$Sr$^+$ is shown in \figref{levelDiagram}.  Dopper cooling is performed on the 422 nm $S_{1/2} \leftrightarrow P_{1/2}$ transition (up to 25 $\mu$W in a 15 $\mu$m radius beam, red detuned by 10--50 MHz from the $^{88}$Sr$^+$ transition) with an additional 1092 nm laser to repump the $D_{3/2}$ state (10 $\mu$W in a 30 $\mu$m radius beam, blue detuned by about 10 MHz from the $^{88}$Sr$^+$ transition).  Resolved sideband cooling and temperature measurement is performed on the 674 nm $S_{1/2} \leftrightarrow D_{5/2}$ transition (800 $\mu$W in a 30 $\mu$m radius beam).  A 1033 nm laser is used to repump the $D_{5/2}$ state between experiments (10 $\mu$W in a 30 $\mu$m radius beam, on resonance with the $^{88}$Sr$^+$ transition).  In the room temperature test setup, the 460 and 405 nm lasers are parallel to the trap axis and the rest of the lasers are at 45 deg to the trap axis.  In the cryogenic test setup, the optical access is more restricted so all of the lasers are parallel to the trap axis with an additional 422 nm laser at 90 deg to the trap axis.

A typical set of voltages for the MIT experiments with $^{88}$Sr$^+$ are given in \tabref{MITvoltages}.  With this set of voltages, the principal axis coordinate system is rotated 14 degrees about the z-axis with respect to the coordinate system shown in \figref{layout}.  The calculated secular frequencies and trap depth are $(\omega_x', \omega_y', \omega_z) / (2 \pi) = (1.7, 2.1, 0.54)$ MHz and 0.025 eV.

\begin{table}
\centerline{
\begin{tabular}{ll}
Electrode		& Voltage \\
\hline
RF			& 155 V at 40.6 MHz \\
C			& 0.95 V \\
2T			& 6 V \\
3B			& 1.3 V \\
4T			& 6 V
\end{tabular}
}
\tcaption{\label{tab:MITvoltages}
Typical voltages used for trap testing at MIT.  The electrodes not listed are grounded.  These voltages trap the ion between electrodes 3B and 3T.}
\end{table}

\subsection{Room temperature testing}
\label{sec:RTtrapTesting}

For room temperature testing, the trap is mounted in a UHV chamber evacuated to $7 \times 10^{-8}$ Pa.  Ions are trapped in the largest trap region between the electrodes 2B and 2T.  The trap used for room temperature testing failed when one of the DC electrodes shorted to the substrate after about 3 weeks of continuous use.

At room temperature large stray electric fields of the order of (10 V)/(100 $\mu$m) $\sim$ 10$^5$ V/m are observed, indicated by the fact that the trap voltages which compensate the trap experimentally are up to 10 V from the predicted voltages (the trap is compensated by minimizing the time correlation of the scattered Doppler cooling photons with the phase of the RF trapping voltage \cite{berkeland98}).  Further, only up to two ions can be loaded into the trap simultaneously.  In contrast with the $^{111}$Cd$^+$ and $^{25}$Mg$^+$ test results, however, the stray electric fields do not change when the ion is lost and the trap is reloaded.  They drift relatively slowly with a time constant of several hours.  This is probably due to the fact that all of the laser wavelengths used for loading and manipulation of $^{88}$Sr$^+$ are $\ge 405$~nm, which corresponds to a photon energy of 3.06 eV, and the work function of Al is roughly 4.1 eV.  In contrast, both $^{111}$Cd$^+$ and $^{25}$Mg$^+$ require lasers with photon energy $> 4.1$ eV.  

\begin{figure}
\centering
\includegraphics[width=8cm]{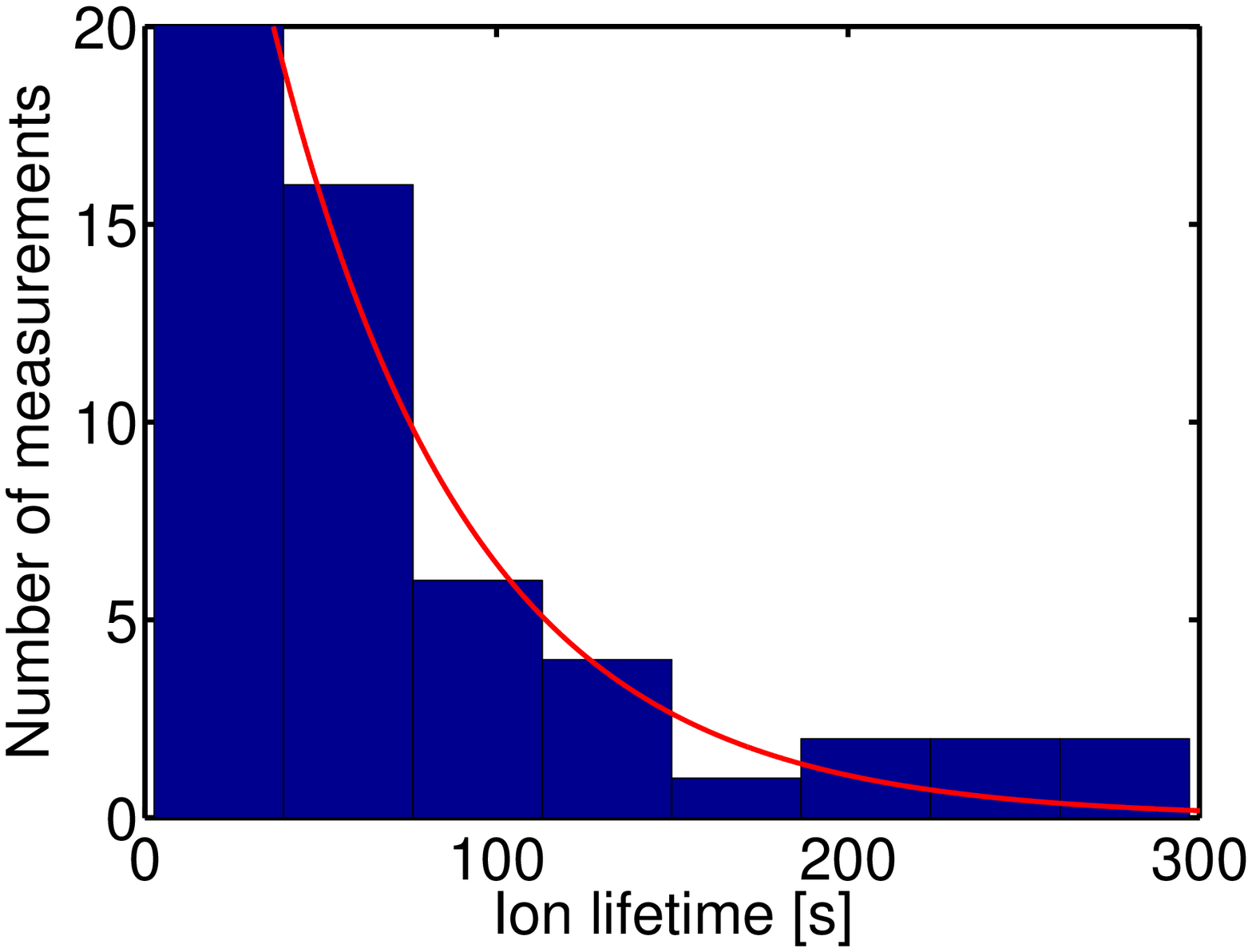} 
\fcaption{\label{fig:RTlifetimeBright}
Histogram of 59 ion lifetimes measured at room temperature with the Doppler cooling lasers on continuously.  An exponential fit with a decay time of $56 \pm 6$ s is overlaid in red.}
\end{figure}

The ion lifetime was measured by loading single ions and measuring how long they remain trapped.  A histogram of 59 ion lifetimes is shown in \figref{RTlifetimeBright}.  In this experiment the Doppler cooling lasers remain on continuously.  The histogram fits well to an exponential with a decay time of $56 \pm 6$ s.  This ion lifetime is not long enough for heating rate measurements.

\begin{figure}
\centering
\includegraphics[width=8cm]{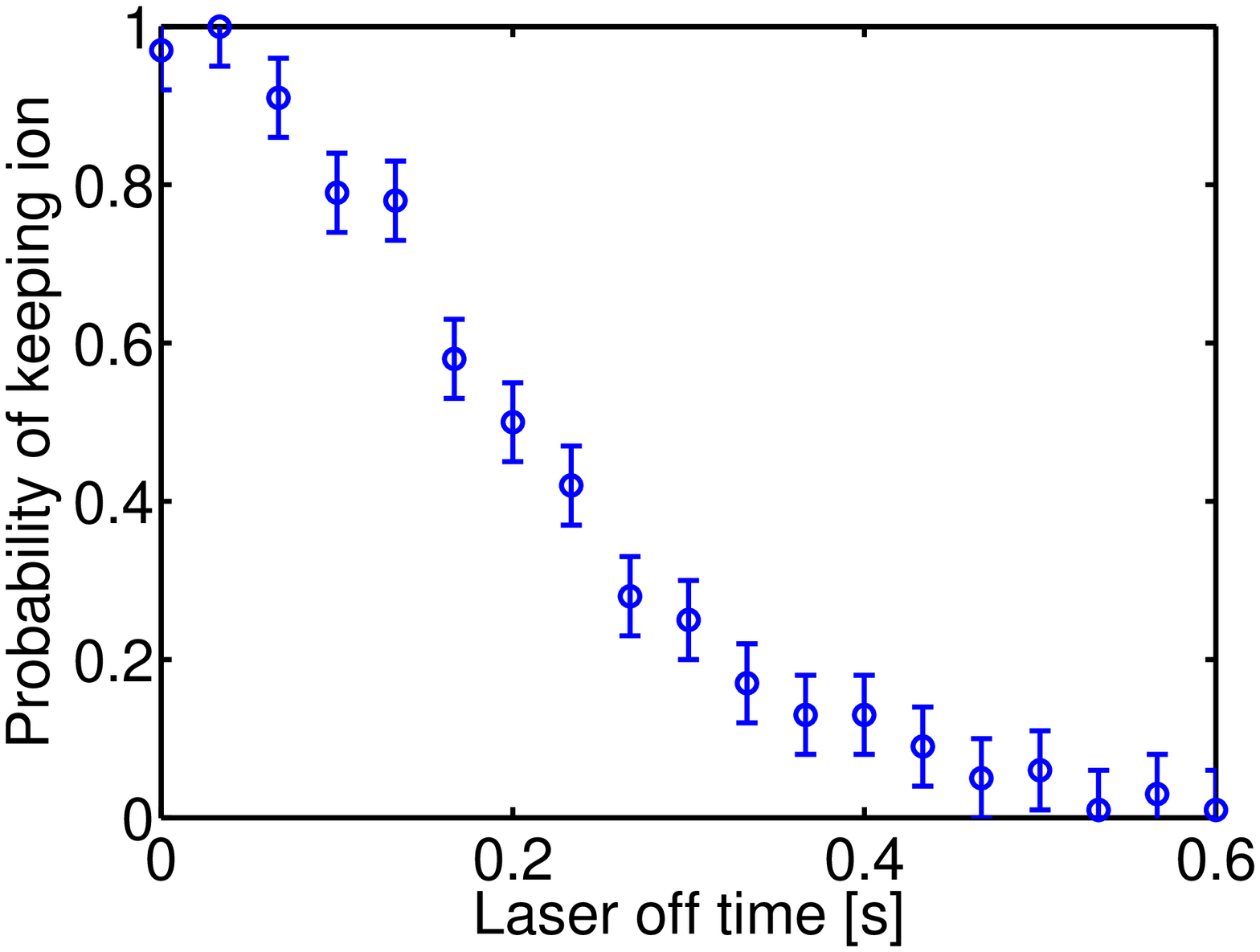} 
\fcaption{\label{fig:RTlifetimeDark}
Probability of keeping an ion as a function of the length of time that the Doppler cooling lasers are turned off, measured at room temperature.  This should fit to an exponential if the ion lifetime is limited by background gas collisions or to a step function if the lifetime is limited by ion heating.  The data has both exponential and step function components, indicating that both background gas and ion heating may play a role in the anomalously short ion lifetime.}
\end{figure}

Measurements of the ion lifetime without Doppler cooling can provide insight into the ion loss mechanism.  The dark lifetime is measured by loading a single ion, turning off the Doppler cooling lasers for a controlled amount of time, and turning the Doppler cooling lasers back on and observing whether the ion is still trapped.  This is repeated many times for each dark time to determine the probability of keeping an ion as a function of the length of time that the Doppler cooling lasers are off, shown in \figref{RTlifetimeDark}.  This should fit to an exponential if the ion lifetime is limited by background gas collisions or to a step function if the lifetime is limited by ion heating.  The data has both exponential and step function components, indicating that both background gas and ion heating may play a role in the anomalously short ion lifetime.  Much longer ion lifetimes have been observed in other surface-electrode ion traps at similar vacuum pressures, so the background gas contribution to the short ion lifetime is not due to the base pressure of the vacuum system.  One possible culprit for the low room temperature ion lifetimes is that the ion lifetime is primarily limited by local outgassing of the SiO$_2$ layer (see \secref{trapDesign}).

\subsection{Cryogenic testing}
\label{sec:cryoTrapTesting}

Recent experiments have shown that cooling ion traps to cryogenic temperatures can improve the heating rate by several orders of magnitude \cite{deslauriers06, labaziewicz07b, labaziewicz08}.  Cooling can also improve the ion lifetime if it is limited by background gas pressure due to local outgassing of the trap itself.

\begin{figure}
\centering
\includegraphics[width=12cm]{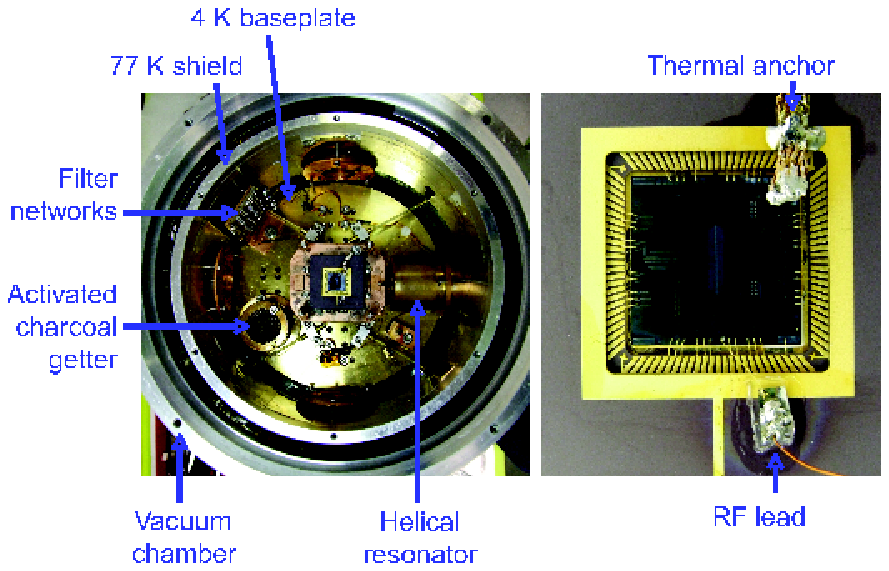}
\fcaption{\label{fig:cryostatPic}
Setup for cryogenic testing.  On the left is a picture of the 4 K baseplate of the cryostat with the trap mounted in the center.  Also visible are the electronic filter networks used to reduce the voltage noise on the DC electrodes (thermally anchored to the 4 K baseplate), the helical resonator used to generate the RF trap drive (thermally anchored to the 77 K shield), and one of the activated charcoal getters used to pump the vacuum chamber (thermally anchored to the 4 K baseplate).  Not pictured are the Sr oven used to load the trap (which is mounted to the front of the helical resonator) and the aspheric lens used to image the ions (which is mounted above the trap inside the vacuum chamber).  On the right is a picture of the trap.  The copper braid in the top right corner is used to provide a good thermal contact between the trap and the 4 K cryostat baseplate.  The RF lead is not routed through the CPGA to reduce capacitance.}
\end{figure}

For cryogenic testing, a second trap was mounted to the 4 K baseplate of a liquid helium bath cryostat \cite{antohi09}.  The cryostat consists of a room temperature vacuum shield, a 77 K thermal shield anchored to a liquid nitrogen dewar, and a 4 K liquid helium dewar.  \figref{cryostatPic} shows the baseplate of the liquid helium dewar.  Optical access is provided by three windows in the plane of the helium baseplate spaced by 90 degrees and one window oriented orthogonal to the helium baseplate.  The cryostat is pumped by a turbopump and by two activated charcoal getters anchored to the 77 K shield and the 4 K baseplate.  While it is difficult to measure the vacuum pressure in the cryostat directly, the partial pressure of oxygen is upper bounded by the ion lifetime measured in other traps to be $\le 3 \times 10^{-10}$ Pa.

Since the CPGA chip carrier is a poor thermal conductor, one end of a copper braid is indium soldered to the silicon substrate of the trap and the other end is mechanically clamped to the 4 K baseplate of the cryostat.  The copper braid provides a strong thermal contact.  The trap is dominantly heated by the power dissipated in the RF electrodes.  The temperature of the RF electrodes is estimated to be of the order of a few tens of K while the temperature of the silicon substrate and DC electrodes is measured in other traps with similar RF drive parameters and capacitance to be 6 K.

Ions are trapped in the largest trap region between the electrodes 3B and 3T.  Stray electric fields of the order of (0.1 V)/(100 $\mu$m) $\sim$ 10$^3$ V/m are observed, and up to 10 ions can be loaded into the trap simultaneously.  The cryogenic testing trap failed when some of the electrodes delaminated from the insulators after temperature cycling between room temperature and 6 K five times and cleaning in water and isopropanol.

\begin{figure}
\centering
\includegraphics[width=6cm]{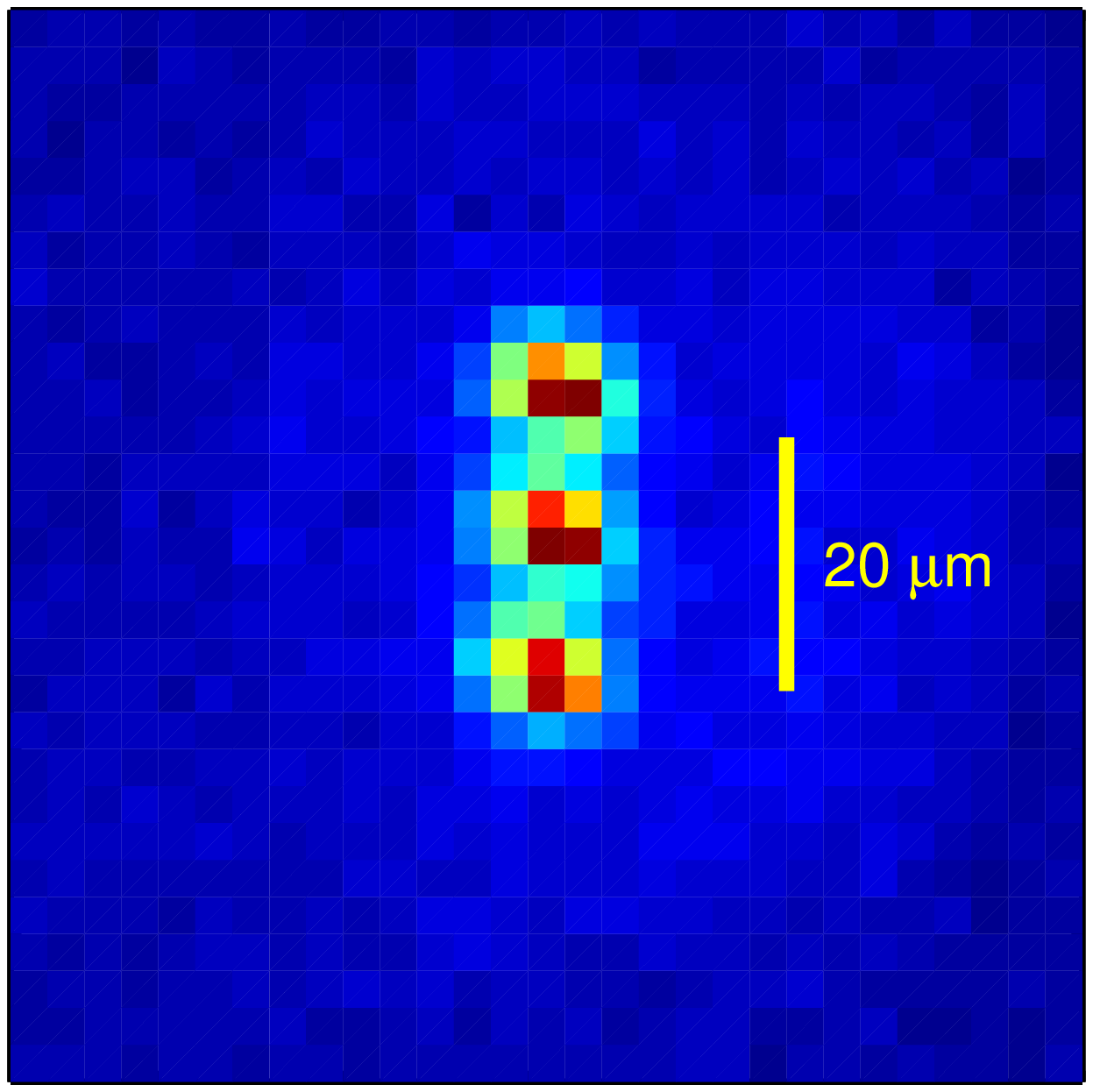} 
\fcaption{\label{fig:ionPic}
An image of three trapped $^{88}$Sr$^+$ ions.  The ion lifetime at 6 K is measured to be $4.5 \pm 1.1$ hours using these ions.}
\end{figure}

The ion lifetime at 6 K was measured using the three ions shown in \figref{ionPic} to be $4.5 \pm 1.1$ hours.  This ion lifetime is improved by two orders of magnitude over the lifetime at room temperature and is long enough for both heating rate measurements and QIP.

\begin{figure}
\centering
\includegraphics[width=8cm]{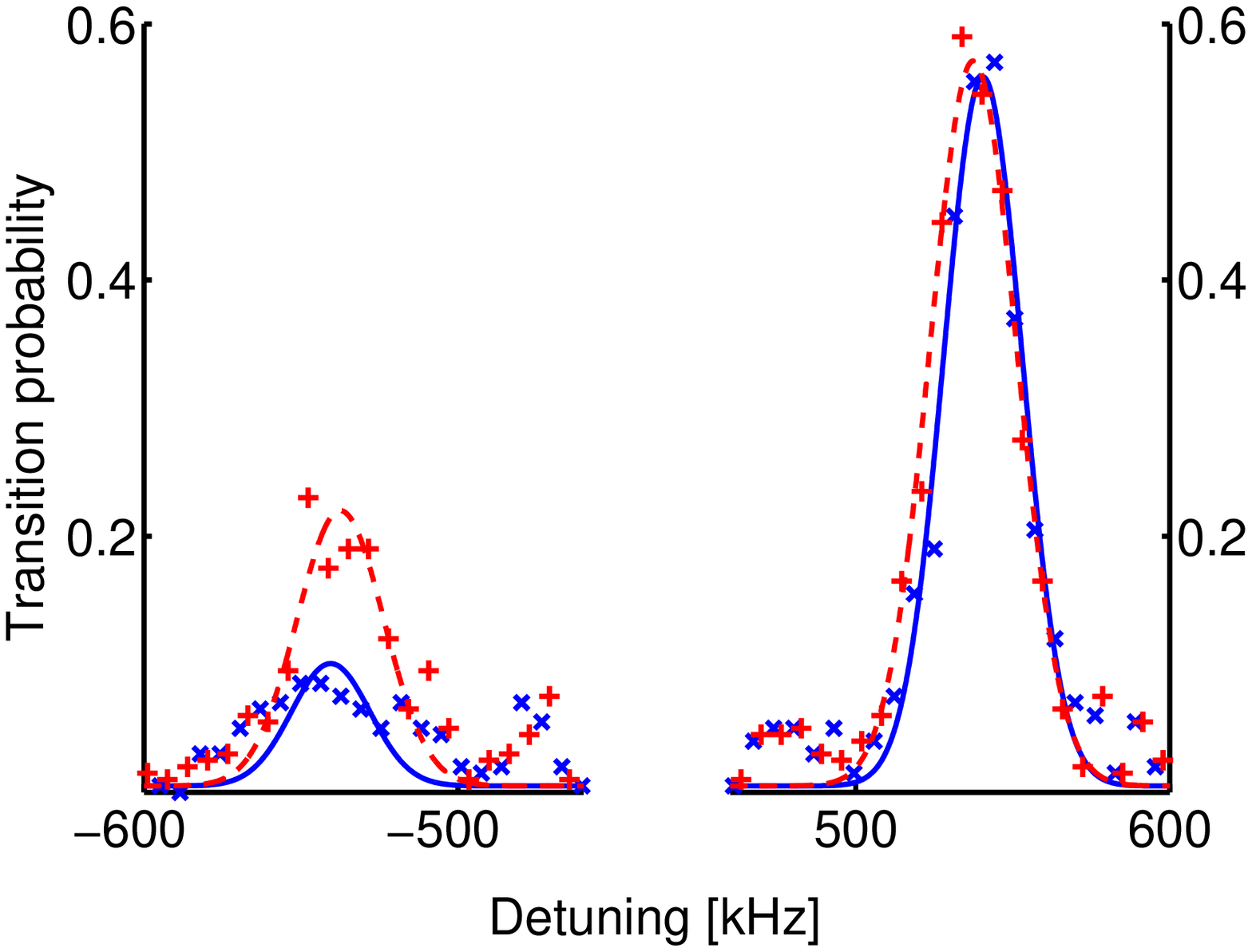} 
\fcaption{\label{fig:sidebandPlot}
The red (left) and blue (right) motional sidebands of the 674 nm $S_{1/2}, m = -1/2 \leftrightarrow D_{5/2}, m = -5/2$.  The blue x's (red +'s) are taken with a 0 ms (2 ms) delay between sideband cooling and measurement.  The blue solid (red dashed) line is a fit used to determine the average number of quanta in the axial motional mode 0 ms (2 ms) after sideband cooling.}
\end{figure}

\begin{figure}
\centering
\includegraphics[width=8cm]{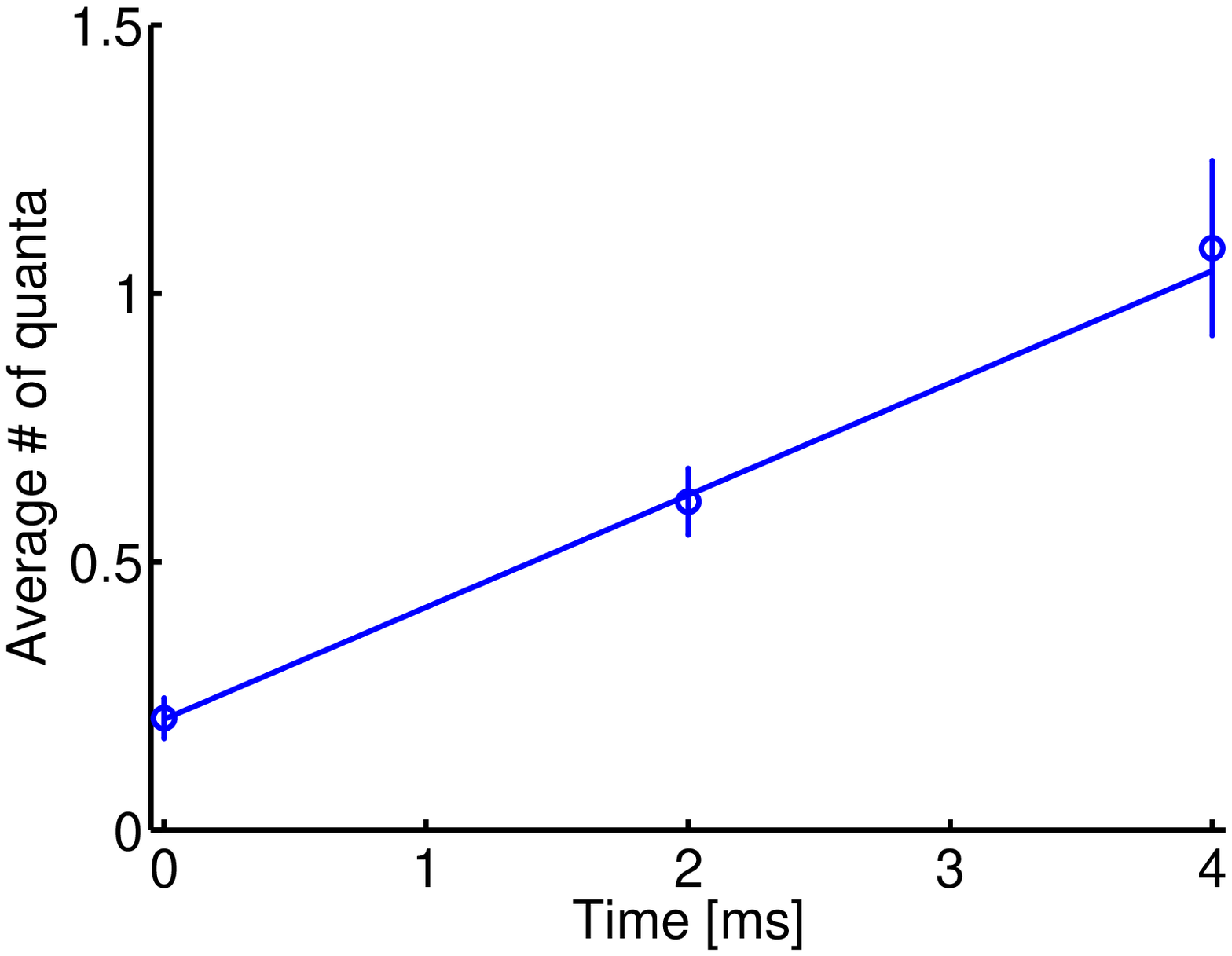} 
\fcaption{\label{fig:nVsTime}
The average number of motional quanta in the axial mode as a function of wait time between sideband cooling and temperature measurement.  The slope of the linear fit (blue line) gives the heating rate.}
\end{figure}

The heating rate is measured by sideband cooling to the ground state of the axial motional mode; waiting 0, 2, or 4 ms; and measuring the relative sizes of the red and blue motional sidebands to determine the expectation value of the number of motional quanta as a function of time.  \figref{sidebandPlot} shows the red and blue motional sidebands after sideband cooling and after a 2 ms wait.  The average number of quanta in the axial motional mode as a function of wait time after sideband cooling is shown in \figref{nVsTime}, along with a linear fit which determines the heating rate in quanta per second.

\begin{figure}
\centering
\includegraphics[width=8cm]{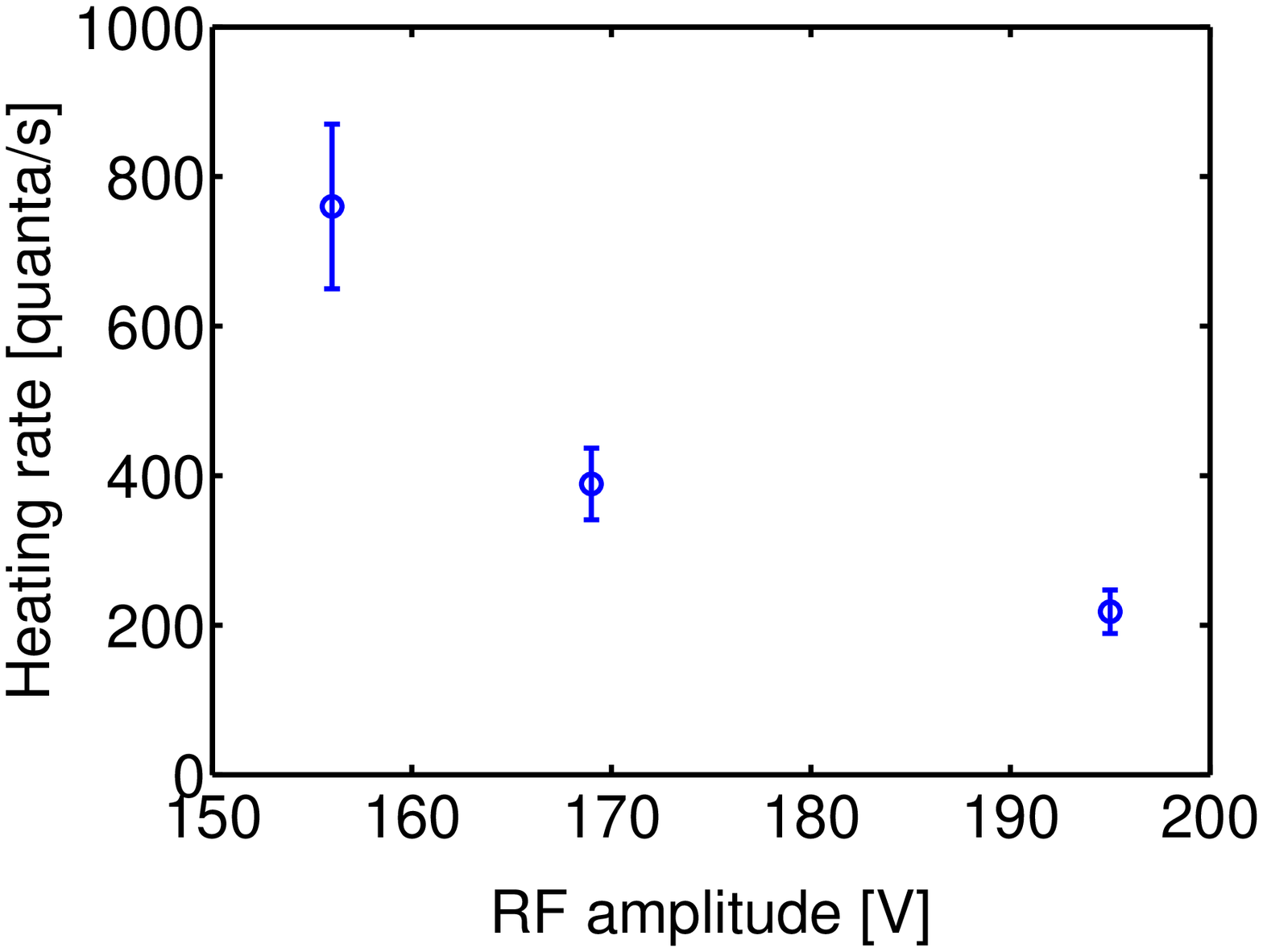} 
\fcaption{\label{fig:heatingRate}
Heating rate of the axial motional mode as a function of the amplitude of the RF trap drive.  The heating rate is smaller at higher RF amplitudes.  This suggests that part of the ion heating measured here is due to coupling of the initially cool axial mode to the relatively hot radial modes.  The radial modes move further away in frequency at higher RF amplitude, which reduces the coupling.  The intrinsic heating rate of the trap at cryogenic temperatures is thus $\le 220 \pm 30$ quanta/s.}
\end{figure}

The observed heating rate, shown in \figref{heatingRate}, is a strong function of the RF amplitude.  This is unexpected because the RF amplitude does not affect the frequency of the axial motional mode.  This is probably due to thermalization of the axial motional mode with the radial modes.  The radial modes are not sideband cooled and the $y'$ radial mode is only weakly Doppler cooled due to the small projection of the Doppler cooling lasers k-vectors on the $y'$ principal axis.  As the RF amplitude is turned up, the radial modes move further away from the axial mode in frequency and the coupling between the radial and axial modes is reduced.  The intrinsic heating rate of the trap is upper bounded by the heating rate measured at the highest RF amplitude: $220 \pm 30$ quanta/s.

\section{Conclusion}
\label{sec:conclusion}

\begin{sidewaystable}
\centerline{
\begin{tabular}{llllll}
Group				& Michigan			& \multicolumn{2}{l}{NIST}						& \multicolumn{2}{l}{MIT}  \\
\hline
Ion species			& $^{111}$Cd$^+$		& \multicolumn{2}{l}{$^{25}$Mg$^+$}				& \multicolumn{2}{l}{$^{88}$Sr$^+$} \\
Photoionization		& 229 nm (pulsed)		& \multicolumn{2}{l}{285 nm}						& \multicolumn{2}{l}{460 and 405 nm} \\
wavelength(s)			&					&					&						&					& \\
Micromotion			& scatter maximization	& \multicolumn{2}{l}{RF sidebands, photon correlation}	& \multicolumn{2}{l}{photon correlation} \\
compensation method	&					&					&						&					& \\
\hline
Trap					& MI I				& NIST I				& NIST II					& MIT I	 			& MIT II \\
Testing temperature		& 300 K				& 300 K				& 300 K					& 300 K				& 6 K \\
RF electrode spacing	& 150 $\mu$m			& 150 $\mu$m			& 125 $\mu$m				& 150 $\mu$m			& 150 $\mu$m \\
Ion height				& 79 $\mu$m			& 79 $\mu$m			& 60 $\mu$m				& 79 $\mu$m			& 79 $\mu$m \\
RF trap drive			& 370 V at 32 MHz		& 140 V at 52 MHz		& 130 V at 51 MHz			& 155 V at 40.6 MHz		& 155 V at 40.6 MHz \\
Secular frequencies		& 5.7, 5.9, 0.5 MHz		& 5.0, 5.4, 1.0 MHz		& 7.3, 7.7, 1.6 MHz			&1.7, 2.1, 0.54 MHz		& 1.7, 2.1, 0.54 MHz \\
Principal axis rotation	& 12 deg				& 11 deg				& 15 deg					& 14 deg				& 14 deg \\
Trap depth [meV]		& 500				& 50					& 64						& 25					& 25 \\
Ground plane present?	& no					& no					& no						& no					& no \\
Loading slot present?	& no					& yes				& yes					& no					& no \\
Attachment method		& epoxy				& ceramic paste		& solder					& solder				& solder \\
Q					& $\sim 100$			& $\sim 90$			& $\sim90$				& $\sim 100$			& $\sim 100$ \\
Vacuum pressure		& $2.0 \times 10^{-8}$ Pa	& $2.0 \times 10^{-8}$ Pa	& $6.7 \times 10^{-9}$ Pa		& $7 \times 10^{-8}$ Pa	& $\le 3 \times 10^{-10}$ Pa \\
Ion lifetime			& 90 s				& 30 -- 120 s			& $300 \pm 30$ s			& $56 \pm 6$ s			& $4.5 \pm 1.1$ hours \\
Heating rate			& ---					& ---					& $7000 \pm 3000$ s$^{-1}$	& ---					& $220 \pm 30$ s$^{-1}$ \\
Spectral noise density	& ---					& ---					& $(8 \pm 3) \times 10^{-11}$	& ---					& $(10 \pm 1) \times 10^{-13}$ \\
at 1 MHz [V$^2$/m$^2$/Hz]
					&					&					&						&					& \\
Trap failure mode		& electrical shorts		& electrical shorts		& did not fail				& electrical shorts		& electrode \\
					& between DC			& between DC			&						& between DC			& delamination \\
					&  and ground			& and ground			&						& and ground			& from the substrate
\end{tabular}
}
\tcaption{\label{tab:summary}
Summary of the trap testing results from all three groups for each of the traps that were successfully used to trap ions at either room temperature or 6 K.}
\end{sidewaystable}

\tabref{summary} summarizes the trap testing.  At room temperature, the traps displayed several problems.  The ion lifetimes measured by all three groups were shorter than expected for the vacuum pressures that were achieved (ion lifetimes between 1 and 5 minutes at vacuum pressures as low as 6.7$\times$10$^{-9}$ Pa).  This is speculated to be due to local outgassing from the trap itself, perhaps in particular the 10 $\mu$m SiO$_2$ layer.  The differences between the ion lifetimes observed by the different groups may be due to some combination of the different trap depths, vacuum bakeout procedures, ion masses, and ion reactivities.  Some of the traps developed shorts between electrodes either during vacuum bakeout or after trapping ions for a few weeks.  The most likely cause of the electrode shorting problem is the thin silicon nitride insulating layer between the electrodes and the ground plane or Si substrate. This layer is especially stressed beneath the wire bonding pads. Shorts due to cracks in the brittle nitride underneath the wire bonding pads or elsewhere in the trap may result in shorts with applied voltages over time. This problem can probably be corrected by using SiO$_2$ instead of the nitride layer since SiO$_2$ is a more robust insulator. The nitride was used for these initial traps because of its relatively high dielectric constant; however, mechanical instability and electrical shorting may outweigh this advantage.  Furthermore, there were also some unidentified reliability issues which prevented some of the traps from working even when the electrodes were not shorted.  Despite these problems, however, the ion heating rate at room temperature was relatively low for a trap of this size: 7 $\pm$ 3 quanta/ms measured with $^{25}$Mg$^{+}$ at a trap frequency of 1.6 MHz and the ion 60 $\mu$m above the trap surface.  This corresponds to a spectral density of electric field noise of $(8 \pm 3) \times 10^{-11}$ V$^2$/m$^2$/Hz at a frequency of 1 MHz (assuming that the spectral density of electric field noise scales like 1/f as observed in other experiments \cite{deslauriers06, labaziewicz07b, labaziewicz08}).

While this trap is unsuitable for QIP at room temperature due to the anomalously short ion lifetime, its behavior is much improved at 6 K.  The stray electric fields are reduced and the ion lifetime is increased by two orders of magnitude to 4.5 $\pm$ 1.1 hours.  The ion heating rate with $^{88}$Sr$^+$ is $\le 220 \pm 30$ quanta/s at a trap frequency of 540 kHz and an ion height of 79 $\mu$m.  This corresponds to a spectral density of electric field noise of $(10 \pm 1) \times 10^{-13}$ V$^2$/m$^2$/Hz at a frequency of 1 MHz, which is suppressed by two orders of magnitude below the room temperature measurement with $^{25}$Mg$^+$ and is only about an order of magnitude higher than the lowest heating rate observed in a trap of this size \cite{labaziewicz08}.  This heating rate is low enough for fault-tolerant QIP operations \cite{steane03}.

Further development is necessary before this trap is ready for large-scale QIP.  The electrode shorting and trapping reliability issues must be addressed.  If the trap is to be used at room temperature, then the ion lifetime must also be improved.  The ion heating rates observed both at room temperature and especially at 6 K, however, are already low enough for high fidelity QIP operations.  Future work will address electrode shorting and trapping reliability, move towards more complicated trap layouts with intersections, and pursue integration of these traps with control electronics and optics.

\nonumsection{Acknowledgements}
\noindent
This work was supported in part by IARPA.  Work at NIST was supported by IARPA and the NIST Quantum Information Program.

\nonumsection{References}


\end{document}